\documentclass[12pt,openbib]{article}
\usepackage{graphicx}
\usepackage{bookman}

\title{\bf Quantum Information Science and Nanotechnology}
\date{}
\author{\em Alexander Yu.\ Vlasov\thanks{%
\protect\includegraphics[scale=0.24]{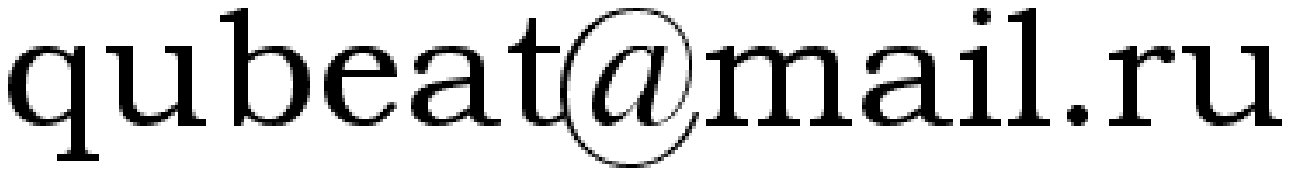}}}

\begin{document}
\maketitle
\begin{abstract}
\noindent
In this note is touched upon an application of quantum information science 
(QIS) in nanotechnology area. The laws of quantum mechanics may be very 
important for nano-scale objects. A problem with simulating of quantum 
systems is well known and quantum computer was initially suggested by 
R.~Feynman just as the way to overcome such difficulties. Mathematical 
methods developed in QIS also may be applied for description of 
nano-devices. Few illustrative examples are mentioned and they may be 
related with so-called fourth generation of nanotechnology products.
\end{abstract}

\section{Brief history}

The beginning of era of nanotechnology is often associated with two 
works: the lecture of R. Feynman \cite{plenty} at Caltech 
in 1959 and book of E. Drexler \cite{eoc} devoted to nanotechnology and 
published in 1986. Both works mentioned examples from (molecular) biology as 
an inspiration. 

Later, the ``pure technical'' approach was criticized sometimes and an 
example is the debate of Drexler and R. Smalley 
\cite{debate}. Often nanotechnology is 
considered \cite{dir,manuf} using both ``top-down'' approach with application 
of more or less traditional ``macroscopic'' technological processes for 
creation and modification of nanostructures and ``bottom-up'' one with 
``shaping things atom by atom''. 

It may be reasonable to consider impact of chaotic motion described 
by thermodynamic laws and additional indeterminacy due to principles 
of quantum mechanics. If the size of few nanometers would be really 
an obstacle for creation of difficult devices, how do biological 
molecular ``machines'' \cite{gene} work? On the other hand, if it 
possible to compare bacteria or virus with ``nanobot'', i.e., 
nano-mechanical device with processor, memory, and different mechanisms 
for motion and interaction with environment? 

An accurate consideration let us avoid some illusory contradictions. 
The thermodynamic problem due to miniaturization were investigated 
in the works of R. Landauer, C. Bennett \cite{heat}, et al quite long 
time ago and should be discussed elsewhere. It is enough to mention, that 
heating may be reduced to zero for some kind of computing devices 
\cite{heat}.

An application of quantum mechanics for description of molecular 
technologies also has certain history. For example, Yu.~I. Manin wrote
about ``genetic automata'' in {\em introduction} to brochure \cite{manin} 
published in 1980: ``Perhaps, for better understanding of this 
phenomenon, we need a mathematical theory of quantum automata''. In 
classical theory of algorithms it is common to use a {\em Turing machine}, 
i.e., an automaton with a tape for program and data. A modification 
of such a model for quantum systems was developed by P. Benioff \cite{ben81} 
on conference PhysComp (Physics and Computations) in 1981. On the 
same conference R. Feynman made talk \cite{sim} affecting problems with 
modeling of quantum systems.

There is specific problem with modeling of nano-systems, because 
description of the system as collection of properties of separate 
atoms already too complicated, but statistical methods may be 
absolutely inappropriate due to certain reasons. 
Modern computers might model evolution of 
billions elementary classical systems. In quantum mechanics the 
situation is quite different due to exponential complication of 
system description with respect to number of elements, and it was 
noted both in work of Yu.~I. Manin cited earlier \cite{manin} and talk of 
R. Feynman mentioned above \cite{sim}. So, even system with three decades of 
particles may require few gigabytes of memory for accurate modeling.

Feynman \cite{sim} suggested creation special simulating device as a method 
to overcome the problem with modeling of the quantum systems. Such a 
device should be universal enough to model as many different quantum 
systems as possible. Feynman introduced term ``quantum computer'' as a 
synonym of ``universal quantum simulator'' and it sometimes causes 
rather limited interpretation of such approach, because the work 
of such device should not necessary be associated with any 
computations.

The Feynman's quantum simulator could be compared with acceleration 
card attached to computer for modeling of quantum systems. D. Deutsch 
\cite{deu85} in 1985 developed further the model of Feynman and 
proved possibility to simulate very extensive class of systems with 
fixed set of elementary operations. The physical motivation of 
considered operations also was emphasized in this work and term 
``algorithm'' might be interpreted not only as computation or computer 
program, but rather as realization of broad class of different 
actions or even as arbitrary physical process.

The book ``Engines of Creation'' by Drexler \cite{eoc} was published in 1986 
and, so, theory of cybernetic devices with atomic scale already 
existed at this time. Drexler had mentioned in his book model of 
Feynman's quantum computers, but nanotechnological applications 
of {\em the quantum information science and technology \sl (QIS\&T)}   
is still at early stage of development.  

\section{Applications}

Let's consider four generations of nanotechnology products, 
cf \cite{manuf}
\begin{itemize}
\item {\bf First Generation.} {\em Passive nanostructures:}
nano-particles, nano-wires, nano-tubes and nano-layers, etc.
\item {\bf Second Generation.} {\em Active nanostructures 
and nanodevices:} transistors, amplifiers, sensors, actuators,
adaptive nano-structures, targeted drugs and chemicals, etc.
\item {\bf Third Generation.} {\em Integrated nanosystems:}
3D nanosystems, multiscale synthesis and assembly 
techniques, bio-assembly, hierarchical structures, etc. 
\item {\bf Fourth Generation.} {\em Heterogeneous multifunctional 
nanosystems:} apparatuses with nanoscale components, where each 
aggregate of atoms has a particular structure and implements its 
own function within the assemblies (``fifth state of matter'').
Radically improved capabilities of this generation may be compared
with some properties of simple natural biosystems still 
not accessible by modern technologies. 
\end{itemize}

QIS\&T is included in ``nanosupplement'' to PACS 2008 (03.67) and
may be associated with all generations of nanotechnologies,
e.g., {\em  quantum dots} belong already to 1'st-2'nd 
generations \cite{manuf} and nowadays are widely used almost
in any area of QIS\&T \cite{intro,qcroad}.

It may be useful to consider application of methods related with 
QIS\&T to basic problem, discussed in 
Drexler-Smalley debate \cite{debate} and related in such a 
classification with fourth generation of nanotechnologies
(and beyond) or with {\em second-generation nanomachnines} 
in Drexler terminology \cite{eoc}. 

Drexler emphasized possibility of creation of so-called 
{\em assemblers}, viz, the ``engines of creation'' \cite{eoc} 
associated with bottom-up approach to
``maneuvering things atom by atom'' \cite{plenty}.
Drexler considered that as the basic purpose of nanotechnology, but 
Smalley (who won the Nobel Prize in Chemistry for discovery of 
fullerenes) did not include conception of the {\em assemblers} in 
nanotechnology and even denied possibility of creation of such 
nano-devices.

Smalley considered idea of some nano-scale manipulator 
for construction of things from separate atoms, as too primitive 
one, because instead of motion in usual space for description of 
molecular interactions it is necessary to use a formal ``many-dimensional 
hyperspace'' \cite{debate}. Indeed, in QIS\&T 
construction of Hamiltonians, providing necessary evolution of a 
system, may be described using geometric algebras for some 
hyperspaces (Clifford algebras) \cite{clif}.

It was already mentioned, the idea of {\em universality} in QIS\&T
\cite{deu85} is concerned not only with possibility of 
realization of any algorithm, but also may be applied to research of 
possibilities and limitation for ``constructing'' of different quantum 
states. Results about universality obtained by Feynman, Deutsch, and 
other researchers ensure preparation of practically {\em arbitrary} 
state using fixed set of simple operations. In 1997 Benioff considered 
mathematical model of quantum robot \cite{ben98} and simplified analogue, 
the ``qubot'' (cf ``nanobot''), was suggested by author of this note 
at ``Think-tank on Computer Science Aspects'' in 2000 \cite{cell}.

The idea of ``quantum processor'' \cite{proc} is also quite useful and may be 
defined as an autonomous system, acting according to internal program. 
It is not necessary some kind of calculations. 
Say, the {\em qubot} may use a program for 
control of motion and interaction with environment.
Similar methods were used in \cite{nondet} for bypassing ``no-go'' results 
like limitations to self-reproducing automata due to laws of 
quantum mechanics formulated by E.~P. Wigner \cite{wig} in 1961  
and ``quantum no-cloning theorem'' by W.~K. Wootters and 
W.~H. Zurek \cite{noclon} in 1982. It may have direct relation 
with the question about possibility of Drexler's assemblers.

\setlength{\bibindent}{-\labelsep}
\setlength{\labelsep}{2\labelsep}
\renewcommand{\refname}{Bibliography}

\end{document}